\documentclass[twocolumn,preprintnumbers]{revtex4}
\usepackage{amsmath}

\usepackage{graphicx}
\usepackage{dcolumn}
\usepackage{bm}
\usepackage{epsfig}

\usepackage{color}

\begin{document}

\title{Energy Systematics of Low-lying Collective States within \\ the
Framework of the Interacting Vector Boson Model}
\author{H. G. Ganev$^1$, V. P. Garistov$^1$, A. I. Georgieva$^{1,2}$ and J. P.
Draayer$^2$}
\address{$^1$\textit{Institute of Nuclear Research and Nuclear Energy,}\\
\textit{\ Bulgarian Academy of Sciences, Sofia1784, Bulgaria}\\
$^2$\textit{Louisiana State University, Department of Physics and Astronomy,}
\\
\textit{Baton Rouge, Louisiana, 70803-4001 USA}}

\begin{abstract}
In a new application of the algebraic Interacting Vector Boson Model
(IVBM), we exploit the reduction of its $Sp(12,R)$ dynamical symmetry
group to $Sp(4,R) \otimes SO(3)$, which defines basis states with fixed
values of the angular momentum $L$. The relationship of the latter to
$U(6) \subset U(3)\otimes U(2)$, which is the rotational limit of the
model, means the energy distribution of collective states with fixed
angular momentum can be studied. Results for low-lying spectra of
rare-earth nuclei show that the energies of collective positive parity
states with $L=0,2,4,6...$ lie on second order curves with respect to the
number of collective phonons $n$ or vector bosons $N=4n$ out of which the
states are built. The analysis of this behavior leads to insight
regarding the common nature of collective states, tracking vibrational as
well as rotational features.
\smallskip

\begin{center}
\end{center}

\end{abstract}

\maketitle

\section{Introduction}

A theoretical description of experimental data on the low-lying collective
states of even-even nuclei in the rare-earth and actinide regions remains
a problem of special interest in the nuclear structure physics. Typically
data is classified from the perspective of sequences of nuclei
\cite{Casten}, with nuclear characteristics studied as a function of the
number of valence nucleons. Such results show the evolution of nuclear
structure as a function of mass number, usually starting with yrast states
of nuclei within a given shell \cite{clasus}. But modern experimental
techniques have advanced to the point that information is often
available on long sequences of non-yrast in addition to yrast states with
$\mathbf{J} ^{\pi }=0^{+},$ $2^{+},$ $4^{+},$ $6^{+}...$ within a given
nucleus \cite{exp}.  In some cases the data is sufficient to justify the
use of a statistical approach for studying the distribution of such
states, like in shell model studies of the spacing distributions for
fixed-particle-rank interactions \cite{stat}. For example, thirteen
$0^{+}$ states have been identified in $^{174}Hf$ \cite{AApr} while in
$^{168}Er$ five $0^{+}$, twelve $2^{+}$, seven $4^{+}$ and seven $6^{+}$
states are known. Many other such examples can be identified. Overall,
theory has fallen short of being able to reproduce the rich array of
experimental results.

In the more traditional quasi-particle-phonon model \cite{Sol},
the nature of the excited bands depends on the number of phonons
and quasi-particle pairs included in the theory. Algebraic
approaches, like the interacting boson model (IBM) \cite{IBM} are
also quite successful in understanding the behavior of the
collective states, and again an important element in the analysis
is the number of collective bosons used to build the states. In
this regard, symplectic models provide a general framework
\cite{Rowe} for investigating collective excitations in many-body
systems as they allow for a change in the number of ``elementary''
excitations in the collective states. The fact that the symplectic
group is also a dynamical group for the harmonic oscillator, which
underpins the shell model, led to the development of very powerful
symplectic shell models \cite{Rowe,dra} where statistical measures
can be used to truncate the symplectic model spaces so that
microscopic calculations are feasible \cite{DraRo}.

Recently an empirical analysis of the data for energies of low-lying
excited states in  even-even nuclei was reported \cite{garistov03}. In
the analysis, experimental energies for $0^{+}$ excited states in the
spectra of the well-deformed nuclei were classified according to the
number of monopole  bosons, using a simple Hamiltonian for generating a
parabolic type energy spectrum \cite{garistov02}. Here, as in the
empirical investigation cited above, we consider systematics in the
behavior of the energies of sequences of collective states with a fixed
angular momentum $L$ in even-even nuclei. The energy distribution of these
states with respect to the number of ``elementary excitations'' (phonons)
that goes into their construction is best reproduced and interpreted
within the framework of algebraic approaches that involve symplectic
symmetries.

A description of the energies of collective states having specific
$L$ values is related to the choice of the dynamical symmetry
group and its reductions. The phenomenological Interacting Vector
Boson Model (IVBM) \cite{IVBM} has been shown to yield an accurate
description of the low-lying spectra of well-deformed even-even
nuclei. The most general spectrum generating algebra of the model
is the algebra of the $Sp(12,R)$ group \cite {Sp12U6}. In the
rotational limit \cite{IVBMrl} of the model, the reduction of
$Sp(12,R)$ to the $SO(3)$ angular momentum group is carried out
through the compact unitary $U(6)$ subgroup, which defines a
boson-number preserving version of the theory.

In the present paper we introduce the ``symplectic" reduction of
$Sp(12,R)$ to the noncompact direct product $Sp(4,R)\otimes
SO(3)$, which isolates sets of states with given $L$ value
(Section II).  In this reduction, $Sp(4,R)$ can be considered as a
classification group for the basis states of the system
\cite{sp4Rclas}. To complete the labelling of the basis state, we use the
reduction of irreducible representations (irreps) of $Sp(4,R)$
into irreps of the pseudo-spin group $SU(2)$
\cite{Sp2NRbr} (Section III). As a result of the correspondence
between the symplectic and unitary reduction chains and the
relation between the second order Casimir operators of $SU(3)$ and
$SU(2)$, we use the same Hamiltonian and basis as in the
rotational limit of the theory \cite{Sp12U6}. The eignevalues of
the Hamiltonian for states with a given value of $L$ give the
energy distribution as a function of the number of excitations $N$
that are used to build the states. The application of this new
version of the IVBM (Section IV) to five even-even rare-earth
nuclei confirms the empirical analysis given in \cite{garistov03},
and extends it to include in addition to deformed nuclei, some
nearly spherical ones. The analysis of the results outlines common
as well as specific features of the two main types of collective
motion (Conclusions).

\section{Group theoretical background}

We consider $Sp(12,R)$, the group of linear canonical
transformation in a $12$-dimensional phase space \cite{MoQu1}, to
be the dynamical symmetry group of the IVBM \cite{IVBM}. Its
algebra is realized in terms of creation (annihilation) operators
$u_{m}^{+}(\alpha )(u_{m}(\alpha )=(u_{m}^{+}(\alpha ))^{\dagger
}),$ of $two$ types of bosons differing by a ``pseudo-spin''
projection $\alpha =\pm 1/2$ in a $3$-dimensional oscillator
potential $m=0,\pm 1$. Hence the bosons that ``built up'' the
collective excitations in the nuclear system are components of
$SO(3)$ vectors and form a ``pseudo-spin'' doublet of the $U(2)$
group. The bilinear products of the creation and annihilation
operators of the two vector bosons generate the noncompact
symplectic group $Sp(12,R)$:
\[
F_{M}^{L}(\alpha ,\beta )=_{k,m}^{\sum }C_{1k1m}^{LM}u_{k}^{+}(\alpha
)u_{m}^{+}(\beta ),
\]
\begin{equation}
G_{M}^{L}(\alpha ,\beta )=_{k,m}^{\sum }C_{1k1m}^{LM}u_{k}(\alpha
)u_{m}(\beta ),  \label{generators}
\end{equation}
\[
A_{M}^{L}(\alpha ,\beta )=_{k,m}^{\sum }C_{1k1m}^{LM}u_{k}^{+}(\alpha
)u_{m}(\beta ),
\]
where $C_{1k1m}^{LM}$ are the usual Clebsh-Gordon coefficients and
$L=0,1,2$, with $M=-L,-L+1,...,L-1,L$, define the transformation
properties of (\ref{generators}) under rotations. Being a noncompact
group, the representations of $Sp(12,R)$ are infinite dimensional. The
action space of the operators (\ref{generators}) is in general reducible
and the invariant operator $(-1)^{N}$, where
$N=-\sqrt{3}[A^{0}(p,p)+A^{0}(n,n)]$ ($p=1/2,n=-1/2$)
counts the total number of bosons, decomposes it into even
\textrm{H}$_{+}$ with $N=0,2,4,...,$ and odd \textrm{H}$_{-}$ with $
N=1,3,5,...,$ subspaces.

\subsection{Reduction through the compact $U(6)$}
The operators $A_{M}^{L}(\alpha ,\beta )$ from (\ref{generators})
generate the maximal compact subgroup $U(6)$ of $Sp(12,R)$. So the
even and odd unitary irreducible representations (UIR) of
$Sp(12,R)$ split into a countless number of symmetric UIR of
$U(6)$ of the type $[N]_{6} =  [N,0^{5}] \equiv [N]$, where
$N=0,2,4,...$ for even $N$ values and $N=1,3,5,...$ for the odd
$N$ values \cite{Sp2NRbr}. The rotational limit \cite{IVBMrl} of
the model is further defined by the chain:
\begin{equation}
\begin{tabular}{lllllllll}
$U(6)$ & $\supset $ & $SU(3)$ & $\times $ & $~~U(2)$ & $\supset $ &
$SO(3)$ & $
\times $ & $U(1)$ \\
\multicolumn{1}{c}{$\lbrack N]$} &  & $(\lambda ,\mu )$ &  & $(N,T)$ & $K$ &
\multicolumn{1}{c}{$\ L$} &  & \multicolumn{1}{c}{$T_{0}$}
\end{tabular}
\label{chain}
\end{equation}
where the labels below the subgroups are the quantum numbers
corresponding to their irreps. In this limit, the Hamiltonian is
expressed in terms of the first and second order invariant
operators of different subgroups in (\ref{chain}). The
\textit{complete} spectrum of the system is calculated through the
diagonalization of the Hamiltonian in the subspaces of \textit{all
}the UIR of $U(6)$ belonging to a given UIR of $Sp(12,R)$
\cite{Sp12U6}. Since the reduction from $U(6)$ to $SO(3)$ is
carried out via the mutually complementary groups $SU(3)$ and
$U(2)$ \cite{Qucom}, their Casimir operators
\cite{Sp2NRbr,Mosh,MQcg} as well as their quantum numbers are
related to one another: $T=\frac{\lambda }{2},N=2\mu +\lambda$.
Making use of the latter, the model Hamiltonian can be expressed
as a linear combination of the first and second order Casimir
invariants of the subgroups in (\ref{chain}):
\begin{equation}
H=aN+bN^{2}+\alpha _{3}T^{2}+\beta _{3}L^{2}+\alpha _{1}T_{0}^{2}.
\label{Halmitonian}
\end{equation}
This Hamiltonian is obviously diagonal in the basis:
\begin{equation}
\mid [N]_{6};(\lambda ,\mu );K,L,M;T_{0}\rangle \equiv \mid
(N,T);K,L,M;T_{0}\rangle.
\label{basis}
\end{equation}

\subsection{The reduction through the noncompact $Sp(4,R)$}

In the present application of IVBM we make use of another possible
reduction \cite{MoQu1} of the $Sp(12,R)$ group; namely, through its
noncompact subgroup $Sp(4,R)$:
\begin{equation}
Sp(12,R)\supset Sp(4,R)\otimes SO(3).\   \label{nred}
\end{equation}
The generators of $Sp(4,R)$ are obtained from the vector addition
to $ L=0 $ (scalar products) of the different pairs of vector
bosons $ u_{m}^{+}(\alpha ),(u_{m}(\alpha ))$ $m=0,\pm 1$
representing the $Sp(12,R)$ generators (\ref{generators}).
By construction, all of these operators are rotational scalars and
as a consequence the generators of
$Sp(4,R)$ [$F^{0}(\alpha ,\beta ),G^{0}(\alpha ,\beta )$ and
$A^{0}(\alpha ,\beta)$]  commute with components of the
angular momentum, $\ L_{M}=-\sqrt{2}\sum_{\alpha }A_{M}^{1}(\alpha
,\alpha )$, that generate the $SO(3)$ group; that is, a direct product
of the two groups is realized (\ref{nred}). It follows from this
that the quantum number $L$ of the angular momentum group $SO(3)$
can be used to characterize the representations of $Sp(4,R)$.

The general reduction scheme of the boson representations of  $
Sp(4k,R)$, $k=1,2,...$, to its maximal compact $U(2k)$ and
noncompact $U(k,k)$ subgroups is given in detail in
\cite{Sp2NRbr}. The case of $k=3$, corresponding to the reduction
of $Sp(12,R)$ through the compact $U(6)$ subgroup, was presented
in the symplectic extension of the $U(6)$  number conserving
version of the IVBM \cite{Sp12U6}. The $Sp(4,R)$  represents the
simplest nontrivial $k=1$ case.  The subgroup associated with its
compact content is $U(2)$. It is generated by the Weyl generators
$A^{0}(\alpha ,\beta )$:
\begin{equation}
\begin{tabular}{ll}
$A^{0}(p,n)=\sqrt{\frac{2}{3}}T_{+}$, &
$A^{0}(n,p)=-\sqrt{\frac{2}{3}}T_{-}$,
\\
$A^{0}(p,p)=\sqrt{\frac{1}{3}}N_{+}$, &
$A^{0}(n,n)=\sqrt{\frac{1}{3}}N_{-}$,
\end{tabular}
\label{WGen}
\end{equation}
where $N_{+}(N_{-})$ count the number of particles of each kind.
It is well known that $Sp(4,R)$ is very convenient for the
classification \cite{sp4Rclas} of nuclear properties with respect
to the reduction operators $N = N_{+}+N_{-}$ and the third
projection $T_{0}=-\sqrt{\frac{3}{2}}[A^{0}(p,p)-A^{0}(n,n)]$ of
the pseudo-spin operator $T$. So we can use the equivalent set of
infinitesimal operators that contain in addition to the raising
$T_{+}$ and lowering $T_{-}$ components of the pseudo-spin (see
(\ref{WGen})) the Cartan operators $N$ and $T_{0}$. The operators
$T_{0},T_{\pm }$ close the pseudo-spin algebra $su(2)$. The
operator $N$ generates $U(1)$ and plays the role of the
first-order invariant of $U(2)\supset SU_{T}(2)\otimes U_{N}(1)$.
So this reduces the infinite dimensional representation of
$Sp(4,R)$ to an infinite sum of finite dimensional representations
$[N]_{2} = [N,0] \equiv [N]$ of $U(2)$. The minimum number of
bosons that are required to build a state with given $L$ is
$N_{\min }=L$ for the $L$ even and $N_{\min }=2L$ for the $L$ odd.
The standard labelling of the $SU_{T}(2)$ basis  states is by
means of the eigenvalues $T(T+1)=\frac{N}{2}(\frac{N}{2}+1)$ of
the second  order Casimir operator $\mathbf{T}^{2}$ of $SU(2)$.
Hence $T=\frac{N}{2},\frac{N}{2} -1,...,1$ or $0,$ for each fixed
$N$ in the reduction $Sp(4,R)\supset U(2)$. The other label of the
$SU_{T}(2)$ basis states is provided by the eigenvalues of the
operator $T_{0}$, which are $T_{0}=-T,$ $-T+1,\ldots ,T-1,T$.

In physical applications, the following correspondence between the
two chains ($U(6)$ (\ref{chain}) and $Sp(4,R)$ (\ref{nred})) of
subgroups of $Sp(12,R)$ plays an important role \cite{vanagas}:
\begin{equation}
\begin{tabular}{lllll}
$Sp(12,R)$ & $\supset $ & $Sp(4,R)$ & $\otimes $ & $SO(3)$ \\
$\cup $ &  & $\cup $ &  & $\cap $ \\
$U(6)$ & $\supset $ & $U(2)$ & $\otimes $ & $SU(3)$.
\end{tabular}
\label{correspondence}
\end{equation}
Result (\ref{correspondence}) is a consequence of the equivalence
of the $U(2)$ group of the pseudo-spin $T$ in both chains
($U(6) \supset SU(2) \subset Sp(4,R)$) and its complementarity to
$SU(3)$ in the $SU(3) \subset U(6)$ chain.

\subsection{Representations of $Sp(4,R)$}

As noted above, each $Sp(4,R)$ irrep is infinite dimensional and
consists of a countless number of $U(2)$ irreps. A basis for its
representations in the even \textrm{H}$_{+}$ action space is
generated by a consecutive application of the symmetrically
coupled products of the operators $F^{0}(\alpha ,\beta )$ to the
lowest weight state (LWS) with angular momentum $L$ that labels
the $Sp(4,R)$ irrep under consideration
\cite{vanagas,JEAL00,RoRo}. Each starting $U(2)$ configuration is
characterized by a totally symmetric representation $[L]$ formed
by $ N_{min}=L$ vector bosons. We now give a procedure for
obtaining the rest of the $SU(2)$ irreps in a given $L$ irrep of
$Sp(4,R)$.
To do this, we first partition $r$ into a sum of sets
$(r_1,r_2)$ with both $r_1$ and $r_2$ even and $r_1 + r_2 = r$,
where $r/2$ gives the degree of the $F^{0}(\alpha ,\beta )$ that
is applied to the LWS.  These pairs can be put into one-to-one
correspondence with the irreps of $U(2)$, and upon reduction to
irreps of $SU(2)$ ($[r_{1},r_{2}] \rightarrow [r_{1} -r_{2},0]
\equiv [r_{1}-r_{2}]$), the decompostion is as follows:
\begin{tabular}{lll}
& $\langle r/4\rangle $ &  \\
$[r] = $ & \multicolumn{1}{c}{$\oplus $} & $[r-4i]$, \\
& \multicolumn{1}{c}{$i=0$} &
\end{tabular}
where $\langle r/4 \rangle $ denotes the integer part of the
ratio. Next, the action of the various products of
$F^{0}(\alpha,\beta )$ operators  is given by all inner products
of the representations $[L]$  i.e. $[L]\otimes ([r] \oplus \lbrack
r-4]\oplus ...)$ restricted to two-dimensional Young diagrams.
They are then transformed into $SU(2)$ representations $[k]$
corresponding to $N=N_{min}+r$, $r=0,2,4,6,... $ and
$T=\frac{k}{2}=\frac{N}{2},\frac{N}{2}-1,....,0$ for even $L$
values. The decomposition of $Sp(4,R)$ representations for odd
values of $L=1,3,5$ in the even \textrm{H}$_{+}$ are obtained in
the same way but using the decomposition of $r=2,4,6,...$ into a
sum of pairs of $[r_{1},r_{2}]$ with $r_1$ and $r_2$ both odd
numbers \cite{DGI} and the multiplication starting with even
representations $[L-1]$.  As a result of the multiplication of
each given value of $L$ with an infinite number of $SU(2)$
decompositions of the even numbers $n$, we obtain all the $SU(2)$
irreps contained in the $Sp(4,R)$ representation defined by $L$.
We illustrate this technique for the cases $L=0$ and $L=2,4$ in
the  TABLES \ref{L0}, \ref{L2}, and \ref{L4}. The columns are
defined by the pseudo-spin quantum  number $T=k/2$ and the rows by
the eigenvalues of $N=k_{\max }=L+r$ for $L$ even and $N=k_{\max
}+2=2L+r$ for $L$ odd and $r=0,2,4,6,...$. TABLE \ref{L0} for the
$L=0$ states actually coincides with the decomposition of the even
numbers $r$.

The correspondence (\ref{correspondence}) between the two
$Sp(12,R)$ subgroup chains, together with the relationships
between the quantum numbers of $U(2)$ and $SU(3)$, allows one to
identify the $SU(3)$ irreps $(\lambda =k,$\ $\mu =(N-k)/2)$ that
are shown in TABLES \ref{L0}, \ref{L2}, and \ref{L4}. For a given
value of $N$, these can be compared to the classification scheme
for $SU(3)$ irreps in the even $U(6)$ irreps of $Sp(12,R)$ given
in TABLE I of Ref. \cite{Sp12U6}. The  $SU(2)$ irreps that are
missing from the tables do not contain states with the $L$ value
being considered, the latter being determined by the reduction
rules for $SU(3) \supset SO(3)$ \cite{IVBMrl}.  Except for the
$L=0$ case in TABLE \ref{L0}, in the decomposition of $Sp(4,R)$
representations $L$ into $[k]$ irreps of $SU(2)$, there is a
multiplicity in the appearance of some of the irreps. The symbol
($\rho  ~\times $), where $\rho $ is an integer number, in TABLES
\ref{L2} and \ref{L4} that shows how many times ($\rho $) the
respective irrep $[k]$ appears for the specified $N$ value. This
multiplicity is exactly equal to the multiplicity of $L$ in the
reduction of the corresponding $(\lambda ,\mu )$ irrep of $SU(3)$
to $L$ of $SO(3)$ \cite{vanagas}.
\begin{widetext}
\begin{table}[th]
\caption{$L=0$}
\begin{tabular}{llllll|l|l||}
\cline{2-8} ... & $T=5$ & \multicolumn{1}{||l}{$T=4$} &
\multicolumn{1}{||l}{$T=3$} & \multicolumn{1}{||l}{$T=2$} &
\multicolumn{1}{||l|}{$T=1$} &
\multicolumn{1}{||l|}{$T=0$} & \multicolumn{1}{||l||}{$T/N$} \\
\cline{2-8}\cline{7-7} &  &  &  &  &  & $[0](0,0)$ & $N=0$ \\
\cline{6-8} &  &  &  &  & \multicolumn{1}{|l|}{$[2](2,0)$} &  &
$N=2$ \\ \cline{5-8}
&  &  &  & \multicolumn{1}{|l}{$[4](4,0)$} & \multicolumn{1}{|l|}{} & $%
[0](0,2)$ & $N=4$ \\ \cline{4-8} &  &  &
\multicolumn{1}{|l}{$[6](6,0)$} & \multicolumn{1}{|l}{} &
\multicolumn{1}{|l|}{$[2](2,2)$} &  & $N=6$ \\ \cline{3-8} & ... &
\multicolumn{1}{|l}{$[8](8,0)$} & \multicolumn{1}{|l}{} &
\multicolumn{1}{|l}{$[4](4,2)$} & \multicolumn{1}{|l|}{} &
$[0](0,4)$ & $N=8$
\\ \cline{2-8}
& \multicolumn{1}{|l}{$[10](10,0)$} & \multicolumn{1}{|l}{} &
\multicolumn{1}{|l}{$[6](6,2)$} & \multicolumn{1}{|l}{} &
\multicolumn{1}{|l|}{$[2](2,4)$} &  & $N=10$ \\ \cline{2-8} &
\multicolumn{1}{|l}{...} & \multicolumn{1}{|l}{...} &
\multicolumn{1}{|l}{
...} & \multicolumn{1}{|l}{...} & \multicolumn{1}{|l|}{...} & ... & $...$%
\end{tabular}
\label{L0}
\end{table}
\begin{table}[th]
\caption{$L=2$} \label{L2}
\begin{tabular}{lllll|l|l||l||}
\cline{2-8} ... & $T=5$ & \multicolumn{1}{||l}{$T=4$} &
\multicolumn{1}{||l}{$T=3$} & \multicolumn{1}{||l|}{$T=2$} &
\multicolumn{1}{||l|}{$T=1$} & \multicolumn{1}{||l||}{$T=0$} &
$T/N$ \\ \cline{2-8} &  &  &  &  & $[2](2,0)$ &  & $N=2$ \\
\cline{5-8}
&  &  &  & \multicolumn{1}{|l|}{$[4](4,0)$} & $[2](2,1)$ & $%
[0](0,2)$ & $N=4$ \\ \cline{4-8}
&  &  & \multicolumn{1}{|l}{$[6](6,0)$} & \multicolumn{1}{|l|}{$%
[4](4,1)$} & 2$\times [2](2,2)$ & $$ & $N=6$ \\
\cline{3-8}\cline{4-4} &  & \multicolumn{1}{|l}{$[8](8,0)$} &
\multicolumn{1}{|l}{$[6](6,1)$} & \multicolumn{1}{|l|}{2$\times
[4](4,2)$} & 2$\times [2](2,3)$ & $ [0](0,4)$ & $N=8$
\\ \cline{2-8} & \multicolumn{1}{|l}{$[10](10,0)$} &
\multicolumn{1}{|l}{$[8](8,1)$} & \multicolumn{1}{|l}{2$\times
[6](6,2)$} & \multicolumn{1}{|l|}{2$
\times [4](4,3)$} & 2$\times [2](2,4)$ & $$ & $N=10$ \\
\hline \multicolumn{1}{|l}{...} & \multicolumn{1}{|l}{...} &
\multicolumn{1}{|l}{... } & \multicolumn{1}{|l}{...} &
\multicolumn{1}{|l|}{....} & ... & ... & ...
\end{tabular}
\end{table}

\begin{table}[th]
\caption{$L=4$} \label{L4}
\begin{tabular}{llll|l|ll||l||}
\cline{2-8} ... & $T=5$ & \multicolumn{1}{||l}{$T=4$} &
\multicolumn{1}{||l|}{$T=3$} & \multicolumn{1}{||l|}{$T=2$} &
\multicolumn{1}{||l|}{$T=1$} & \multicolumn{1}{||l||}{$T=0$} &
$T/N$ \\ \cline{2-8}
&  &  &  & \multicolumn{1}{|l|}{$[4](4,0)$} &  &  & $N=4$ \\
\cline{4-5}\cline{8-8}
&  &  & \multicolumn{1}{|l|}{$[6](6,0)$} & \multicolumn{1}{|l|}{$%
[4](4,1)$} &  &  & $N=6$ \\ \cline{3-8}\cline{4-4} &  &
\multicolumn{1}{|l}{$[8](8,0)$} & \multicolumn{1}{|l|}{$[6](6,1)$}
& \multicolumn{1}{|l|}{2$\times [4](4,2)$} & \multicolumn{1}{|l|}{2$%
\times [2](2,3)$} & \multicolumn{1}{|l||}{$[0](0,4)$} & $N=8$ \\
\cline{2-8} & \multicolumn{1}{|l}{$[10](10,0)$} &
\multicolumn{1}{|l}{$[8](8,1)$}
& \multicolumn{1}{|l|}{2$\times [6](6,2)$} & \multicolumn{1}{|l|}{2$%
\times [4](4,3)$} & 2$\times [2](2,4)$ & \multicolumn{1}{|l||}{%
$$
} & $N=10$ \\ \hline \multicolumn{1}{|l}{...} &
\multicolumn{1}{|l}{...} & \multicolumn{1}{|l}{... } &
\multicolumn{1}{|l|}{...} & \multicolumn{1}{|l|}{....} & ... &
\multicolumn{1}{|l||}{...} & ...
\end{tabular}
\end{table}
\end{widetext}

\section{Energy distribution of low-lying collective states}
\subsection{Application of the theory}

Because of the correspondence between the symplectic and unitary
reduction chains (\ref{correspondence}) and the relation between the
$SU(3)$ and $SU(2)$ second order Casimir operators, in the present case
we can use the same Hamiltonian (\ref{Halmitonian}) as in \cite{Sp12U6}.
Furthermore, as established above, the bases in the two cases are
equivalent and as a result the Hamiltonian (\ref{Halmitonian}) is
diagonal in both (\ref{basis}). The eigenvalues for states of a given
$L$ value are therefore simply
\begin{align}
E((N,T);KLM;T_{0})=&aN+bN^{2}+\alpha _{3}T(T+1)+\alpha
_{1}T_{0}^{2}
\nonumber \\
&+\beta _{3}L(L+1).  \label{energies}
\end{align}
In this expression (\ref{energies}) the dependence of the energies of the
collective states on the number of phonons (vector bosons) $N$ is parabolic.
All the rest of the quantum numbers $T$, $T_{0},$ and
$L$ defining the states  are expressed in terms of $N$ by means
of the reduction procedure described
above. This result affirms the conclusions of an empirical investigation
of the behavior of states with fixed angular momentum \cite{garistov03},
namely, that their energies are well described by the simple
phenomenological formula $E_{L}(n)=A_{L}n-B_{L}n^{2},$ where $A_{L}>0$
and $B_{L}>0$ are fitting parameters and $n$ is an integer number
corresponding to each of the states with given $L$. The number $n$
labelling each state is related in \cite{garistov02} to the number of
monopole bosons, which are obtained by means of a Holstein-Primakoff
mapping \cite{HP} of pairs of fermions confined to a $j$-orbit with
projection $m$. In further considerations we have established a relation
$N=4n$ between the quantum number $N$ and the number of ideal bosons $n$
introduced in \cite {garistov02}. This relates the $ Sp(12,R)$
phenomenological collective model to a simple microscopic description of
the collective states.

  From this analysis, it should be clear that the IVBM can be used to
investigate energy distributions (\ref{energies}) of low-lying states
with $J^{\pi }=0^{+},2^{+},4^{+}...$ as a function of the number of
bosons. In the applications that follow, we focus on states with $L=0,2,4$. The
basis states have a fixed $T$ value the parity defined as $\pi =(-1)^{T}$
\cite{Sp12U6}. In the columns with fixed $T$, the number of bosons
$N$ changes in
steps of four ($\Delta N=4)$ in the $L=0,1$ cases and steps of two
($\Delta N=2)$
for all the rest (see examples given in TABLES \ref{L0}-\ref{L4}).

In what follows we present a procedure for obtaining the energy
distribution of low-lying collective states in real nuclei. We
start with an evaluation of the inertia parameter, $\beta _{3}$,
that muliplies the $L(L+1)$ term in (\ref{energies}). This is done
by fitting energies of the ground-state-band (GSB) with
$L=0_{1},2_{1},4_{1}$,$6_{1},...$ to the experimental values for
each nucleus. The other collective states in this approach, which
typically are associated with other terms in the eigenvalues of
the Hamiltonian (\ref{energies}), are also influenced by the value
of the inertia parameter. A convenient method for determining the
other parameters of the Hamiltonian (\ref{Halmitonian}) follows:

\begin{itemize}
\item For the $0^{+}$ states we fix $T=0$ and $T_{0}=0$ with
$N=0,4,8,...$, which corresponds to the first column on TABLE \ref{L0}. As
a result, we obtain a simple two-parameter quadratic equation
\begin{equation}
E((N_{0_{i}},0);000;0)=aN_{0_{i}}+bN_{0_{i}}^{2}  \label{0energ}
\end{equation}
for the energy distributions of the $i=1,2,3,...$ experimentally observed
$0_{i}^{+}$ state. The index $i$ denotes an ordering of the states with
increasing energy.  The $E_{0_{1}^{+}}=0$ equation for the ground state
has two solutions, namely, $N_{0_{1}}^{\prime }=0$ and $N_{0_{1}}^{\prime
\prime }=-\frac{a}{b}$, as the parabola is a symmetric curve with respect
to its maximum value at $N_{0_{\max }}=-\frac{a}{2b}$ for $b<0$. The
values of  $N_{0_{i}}$ corresponding to the experimentally observed
$E_{\exp }$ and the values of the parameters $a$ and $b$  are evaluated in
a multi-step $\chi$-squared fitting procedure. Specifically, the fitting is
performed by comparing the energies of various possible sets of values of
$N_{0_{i}}$ attributed to the experimental data for the $0_{i}^{+}$ states.
The set with minimal $\chi ^{2}$ value determines the distribution of the
$0_{i}^{+}$ states energies ($a$ and $b)$ with respect to the number of
bosons $N_{0_{i}}$ out of which each state is built.

\item  For $2^{+}$ states we leave the $a$ and $b$ parameters fixed as
determined above for $0^{+}$ states and introduce a dependence on
the quantum number $T\neq 0$ (even) and $T_{0}=0$ with allowed
values of $N$ at fixed $T$, $N=2T,2T+2,2T+4,...$ (third, fifth,
etc., column in TABLE \ref{L2}). This determines the
$\alpha_{3}T(T+1)$ part of the Hamiltonian eigenvalue
(\ref{energies}). In this case the $\alpha _{3}$ parameter is
again evaluated by means of a $\chi ^{2}$ fitting to values for
the different sets of the allowed $N_{2_{i}}$  attributed to the
experimentally observed $2_{i}^{+}$ states. This term plus the
constant $6\beta _{3}$ fix the distance the $2^{+}$ parabolas are
separated from the $0^{+}$ ones.

\item  For the $4^{+}$ states, we again use fixed $\ T\neq 0$ (even) and keep
the values of  $a$, $b$ and $\alpha _{3},$ as determined for the
$0^{+}$ and $2^{+}$ states, respectively, but choose  $T_{0} \neq
0$ from the allowed set given by $T_{0}=$ $\pm 1,\pm 2,...,\pm T$.
This choice allows one to evaluate the final parameter in the
Hamiltonian, $\alpha _{1}$.  This parameter determines the
distance the parabolas representing the $4_{i}^{+}$ states are
shifted from those previously determined, with the appropriate set
of values of $N_{4_{i}},$ which gives the energy distribution of
the $4^{+}$ states with respect to $N_{4_{i}}$ by means of the
same type of multi-step fitting.

\end{itemize}

The model parameters are fixed with respect to the $0^{+},2^{+}$
and $4^{+},$ as there is usually enough of these states to achieve
good fitting statistics and they are predominantly bandhead
configurations (all the $0^{+}$ and some of the $2^{+}$ and
$4^{+}$ states).  Sets of states with other values of $ L=1,3,6$
or with negative parity ($T$-odd) can be included in the
consideration by determining in a convenient way the values of
$T,T_{0}$ and finding sequences of $N$ corresponding to the
observed experimental energies. Of course, the quantum numbers
$T,T_{0}$ and the sequences of $N$ have to be allowed in the
reduction of the $Sp(4,R)$ representation defined by $L$, to the
$SU(2)$ irreps. As  all the parameters of Hamiltonian are
evaluated from the distribution of $0^{+},2^{+}$ and $4^{+}$
states, only changing the values of $T$ and $T_{0}$ is not enough
to make the curve for the additional set of states with another
$L$ distinguishable. In order to do this, we can introduce a free
additive constant $c_{L}$ to the eigenvalues $E((N,T);KLM;T_{0})$,
which is evaluated in the same way as the other parameters of
(\ref{energies}) in a fit to the experimental values of the
energies of the additional set of $L$ states being considered.

\subsection{ Analysis of the results}
\begin{figure}[th]
\centerline{\hbox{\epsfig{figure=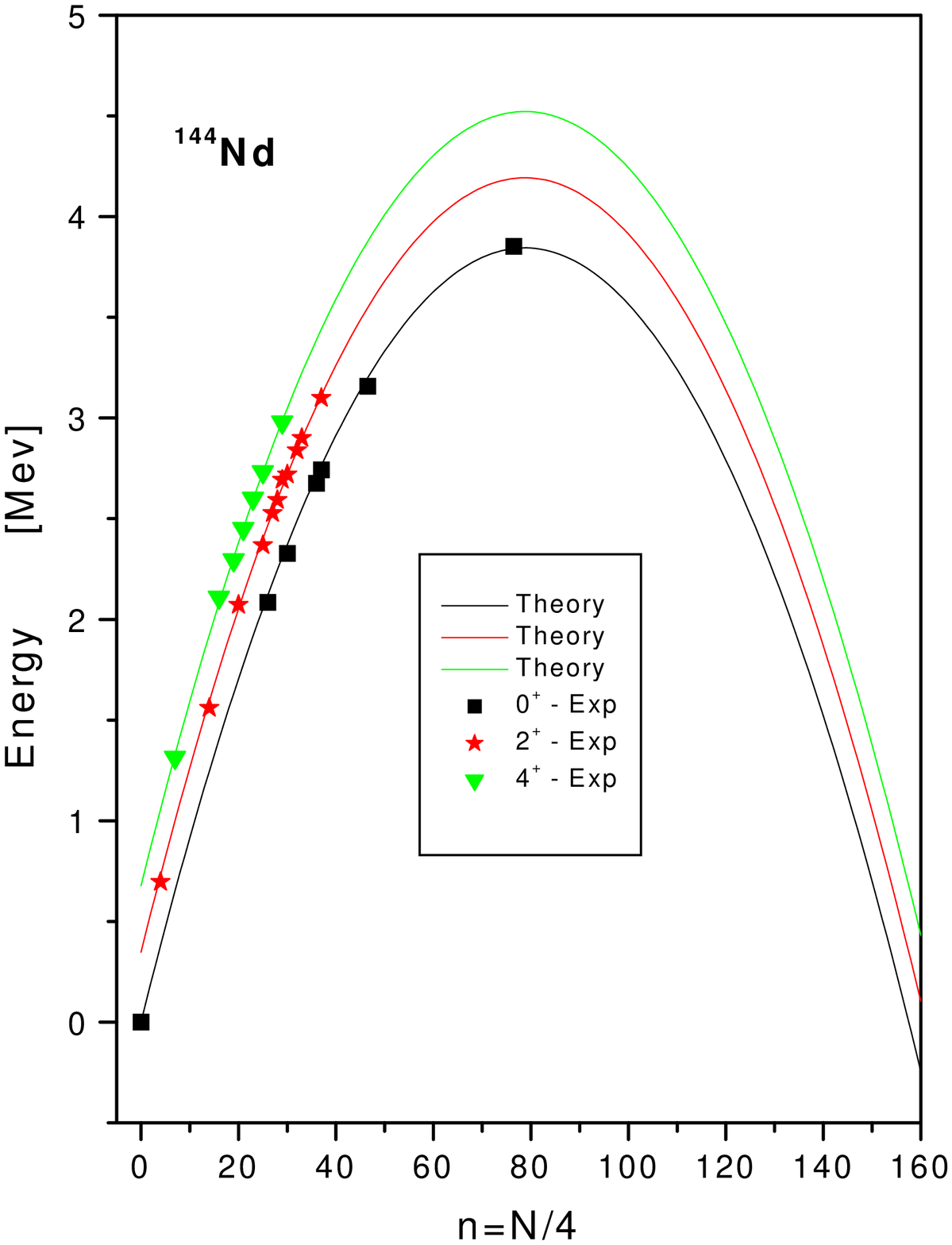,height=6cm}%
\epsfig{figure=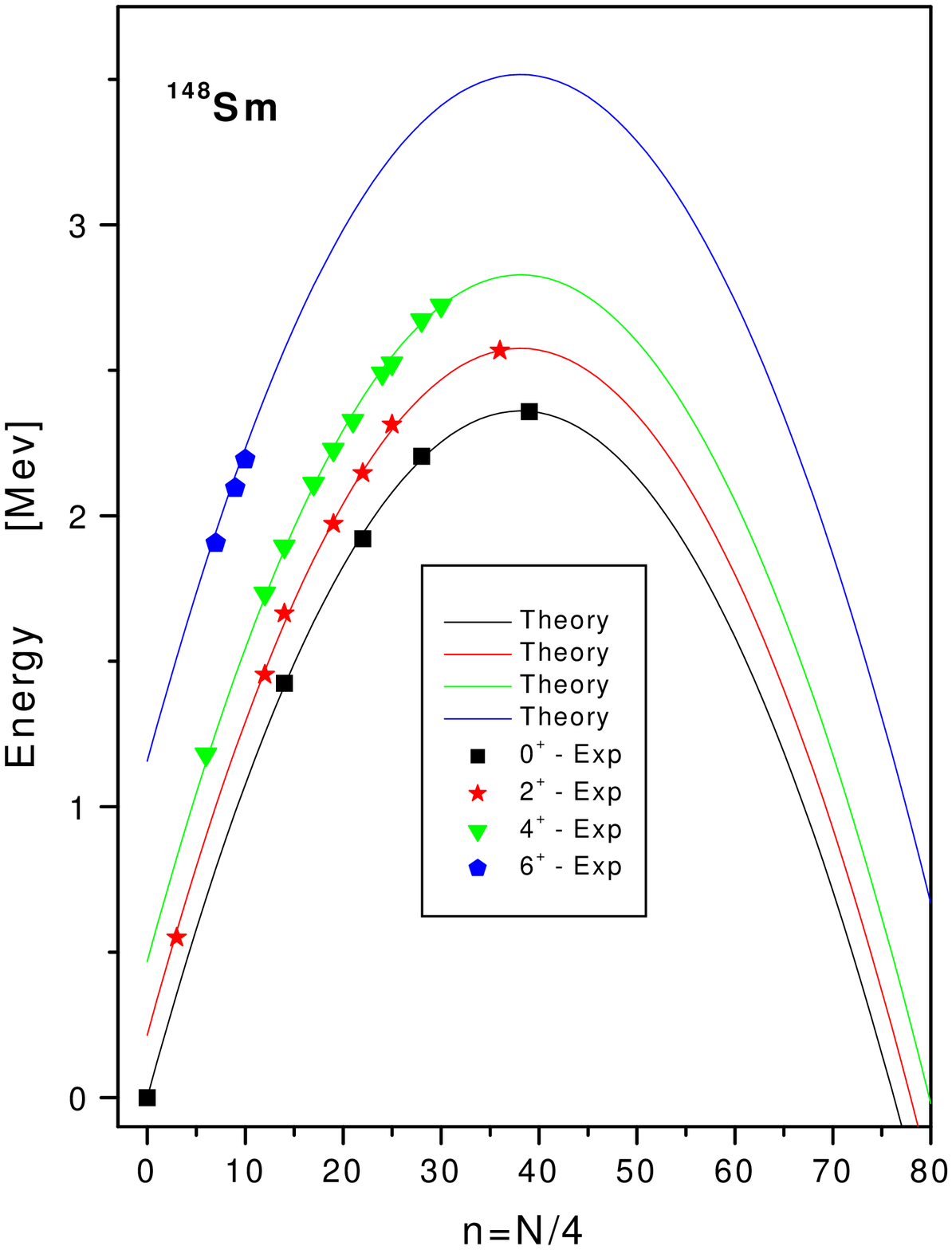,height=6cm}}} \caption{(Color online)
Comparison of the theoretical and experimental energy
distributions of the states with $J^{\pi}=0^{+},2^{+},4^{+}$ for
$^{144}Nd$ (left) and with $J^{\pi}=0^{+},2^{+},4^{+}$ and $6^{+}$
for $^{148}Sm$(right)} \label{NdSm}
\end{figure}

\begin{figure}[th]
\centerline{\hbox{\epsfig{figure=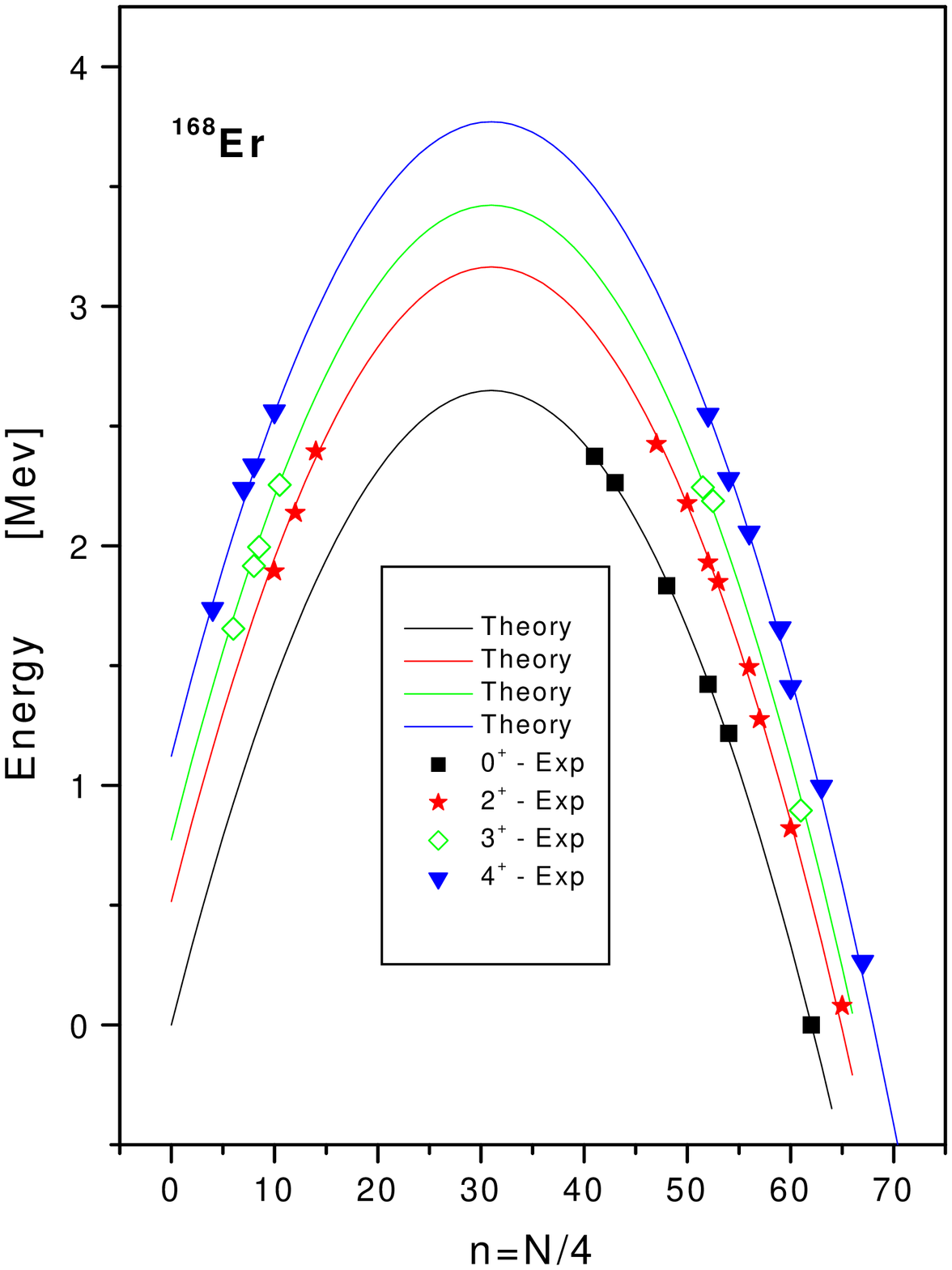,height=6cm}%
\epsfig{figure=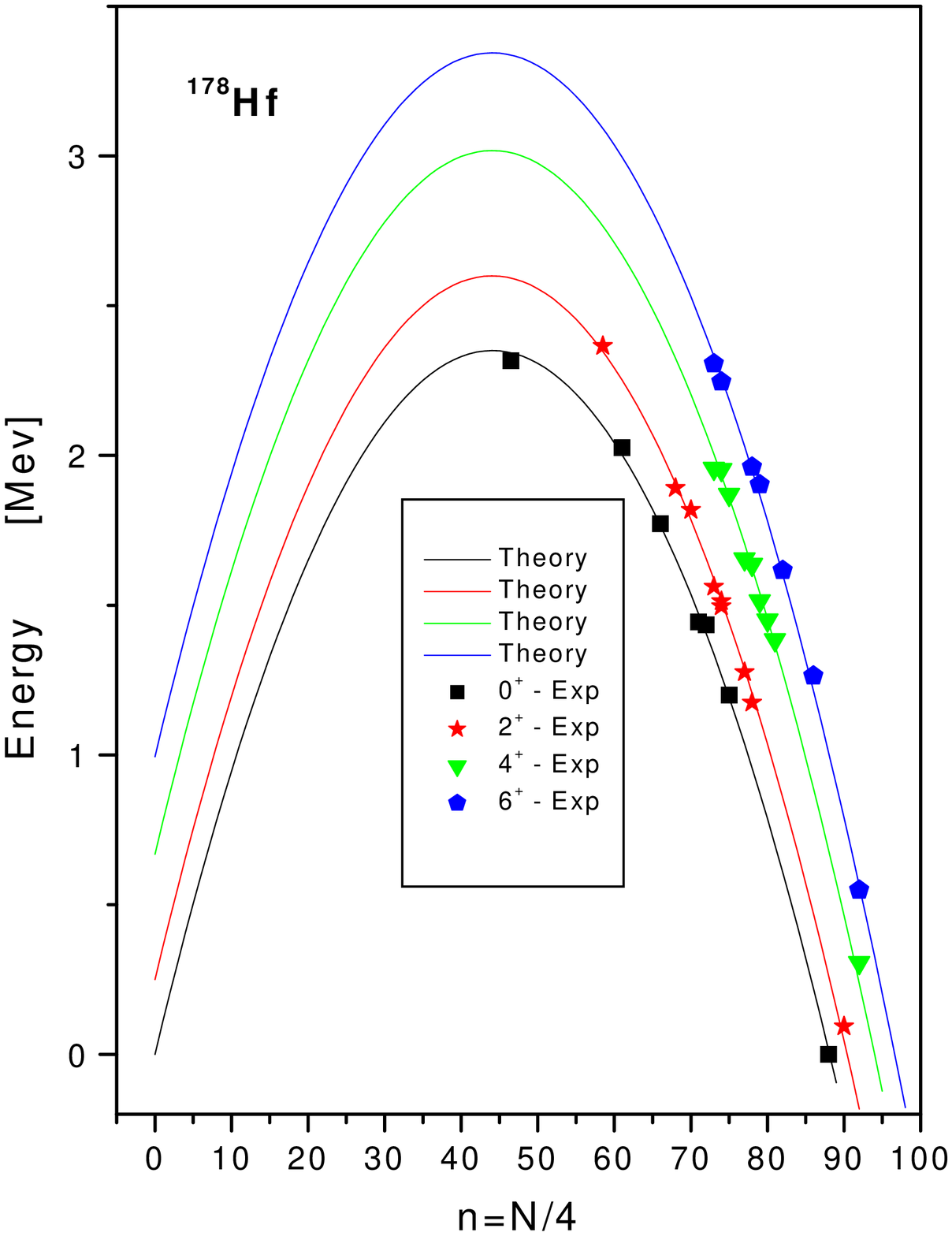,height=6cm}}} \caption{(Color online)
The same as in FIG.1, but for states with
$J^{\pi}=0^{+},2^{+},3^{+}$ $4^{+}$ in $^{168}Er$ (left) and
$J^{\pi}=0^{+},2^{+},4^{+}$ and $6^{+}$ in $^{178}Hf$(right)}
\label{ErHf}
\end{figure}

\begin{figure}[th]
\centerline{\hbox{\epsfig{figure=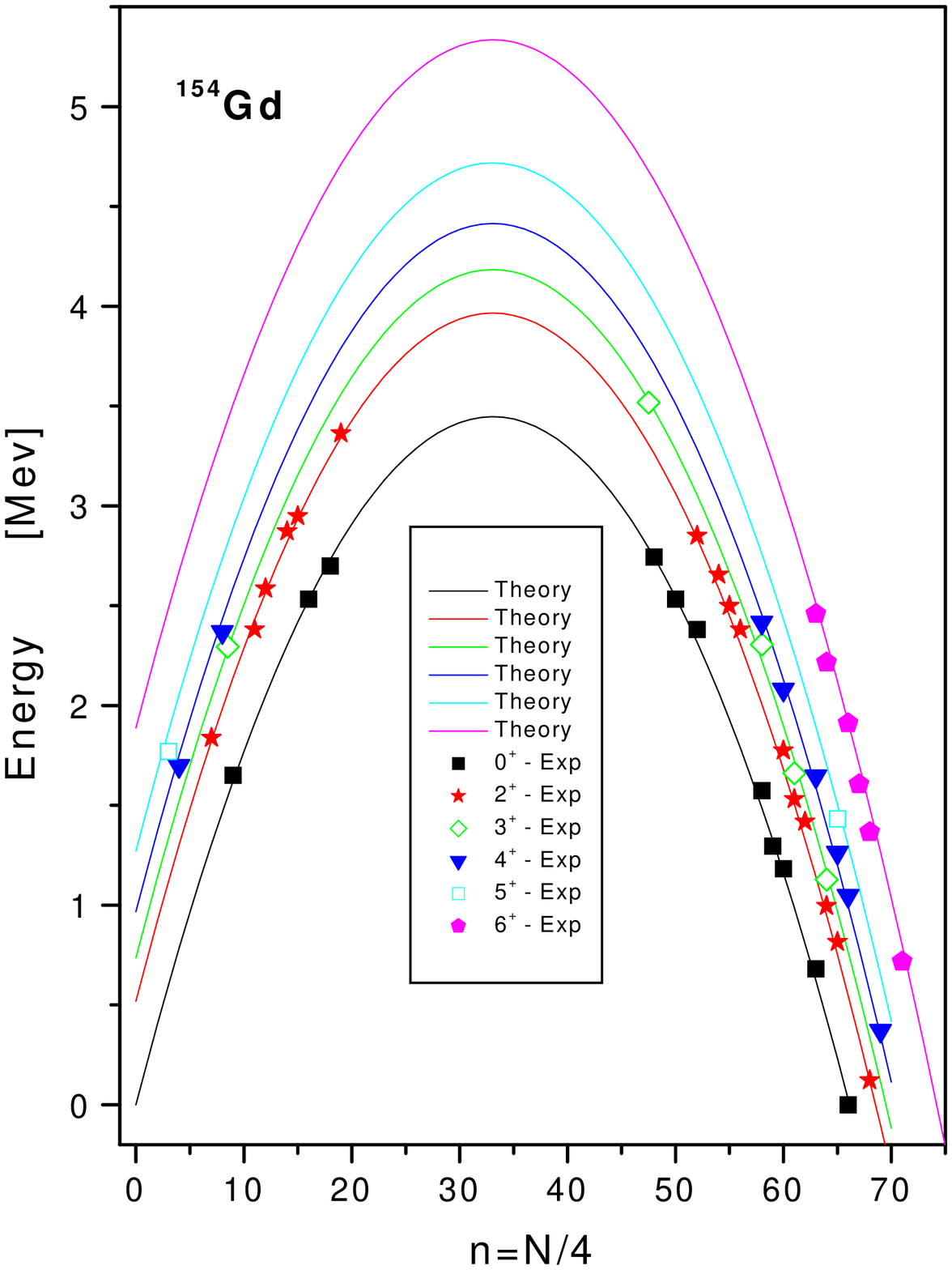,height=7cm}}}
\caption{(Color online) The same as in FIG.1, but for states with
$L=0^{+},2^{+},3^{+},4^{+},5^{+}$ and $6^{+}$ in $^{154}Gd$}
\label{Gd}
\end{figure}

The results of the treatment described above applied to the collective
spectra of five even-even nuclei from the rare earth region are
illustrated in Figures \ref{NdSm}-\ref{Gd}. The distribution of the
energies with respect to values of $N_{L_{i}}$ and the good
agreement between theory and experiment can be clearly seen in the figures.
Additionally $N_{\min}$, the values of $T,T_{0}$ used for the
states with given $L$, the values obtained for the Hamiltonian
parameters $\beta _{3},a,b,\alpha _{3}$, $\alpha _{1}$ and $c_{L}$
with their respective $\chi ^{2}$ are given in TABLE
\ref{coef}. The $s$ in the first column gives the number of the
experimentally observed states with the respective $L$ value.
\begin{table}[th]
\caption{Parameters of the Hamiltonian obtained in the fitting
procedure}
\begin{tabular}{||l||llllll|l||}
\hline\hline Nucleus & $s$ & \multicolumn{1}{||l}{$L$} &
\multicolumn{1}{||l}{$N_{\min }$} & \multicolumn{1}{||l}{$T$} &
\multicolumn{1}{||l}{$T_{0}$} &
\multicolumn{1}{||l|}{$\chi ^{2}$} & \multicolumn{1}{||l||}{parameters} \\
\hline\hline $^{148}Sm$ & 5 & \multicolumn{1}{|l}{0} &
\multicolumn{1}{|l}{0} & \multicolumn{1}{|l}{0} &
\multicolumn{1}{|l}{0} & \multicolumn{1}{|l|}{0.0005 } &
\begin{tabular}{l}
$a=0.03096$ \\
$b=-0.00010$%
\end{tabular}
\\
& 7 & \multicolumn{1}{|l}{2} & \multicolumn{1}{|l}{8} &
\multicolumn{1}{|l}{4 } & \multicolumn{1}{|l}{0} &
\multicolumn{1}{|l|}{0.0002} & $\alpha
_{3}=-0.00187$ \\
& 10 & \multicolumn{1}{|l}{4} & \multicolumn{1}{|l}{6} &
\multicolumn{1}{|l}{8} & \multicolumn{1}{|l}{8} &
\multicolumn{1}{|l|}{0.0003
} & $\alpha _{1}=-0.00285$ \\
& 3 & \multicolumn{1}{|l}{6} & \multicolumn{1}{|l}{20} &
\multicolumn{1}{|l}{ 10} & \multicolumn{1}{|l}{10} &
\multicolumn{1}{|l|}{0.0023} & $\beta _{3}=0.03929$ \\
\hline\hline $^{144}Nd$ & 7 & \multicolumn{1}{|l}{0} &
\multicolumn{1}{|l}{0} & \multicolumn{1}{|l}{0} &
\multicolumn{1}{|l}{0} & \multicolumn{1}{|l|}{0.0001 } &
\begin{tabular}{l}
$a=0.02389$ \\
$b=-0.00003$%
\end{tabular}
\\
& 11 & \multicolumn{1}{|l}{2} & \multicolumn{1}{|l}{12} &
\multicolumn{1}{|l}{6} & \multicolumn{1}{|l}{0} &
\multicolumn{1}{|l|}{0.0008
} & $\alpha _{3}=0.00309$ \\
& 7 & \multicolumn{1}{|l}{4} & \multicolumn{1}{|l}{20} &
\multicolumn{1}{|l}{ 10} & \multicolumn{1}{|l}{10} &
\multicolumn{1}{|l|}{0.0004} & $\alpha
_{1}=-0.00450$ \\
&  & \multicolumn{1}{|l}{} & \multicolumn{1}{|l}{} &
\multicolumn{1}{|l}{} &
\multicolumn{1}{|l}{} & \multicolumn{1}{|l|}{} & $\beta _{3}=0.04074$ \\
\hline\hline $^{168}Er$ & 6 & \multicolumn{1}{|l}{0} &
\multicolumn{1}{|l}{0} & \multicolumn{1}{|l}{0} &
\multicolumn{1}{|l}{0} & \multicolumn{1}{|l|}{0.0006 } &
\begin{tabular}{l}
$a=0.04270$ \\
$b=-0.00017$%
\end{tabular}
\\
& 11 & \multicolumn{1}{|l}{2} & \multicolumn{1}{|l}{4} &
\multicolumn{1}{|l}{ 2} & \multicolumn{1}{|l}{0} &
\multicolumn{1}{|l|}{0.0022} & $\alpha
_{3}=0.07298$ \\
& 11 & \multicolumn{1}{|l}{4} & \multicolumn{1}{|l}{4} &
\multicolumn{1}{|l}{ 2} & \multicolumn{1}{|l}{2} &
\multicolumn{1}{|l|}{0.0015} & $\alpha
_{1}=0.10910$ \\
& 7 & \multicolumn{1}{|l}{3} & \multicolumn{1}{|l}{4} &
\multicolumn{1}{|l}{2
} & \multicolumn{1}{|l}{0} & 0.0009 & $\beta _{3}=0.01295$ \\
&  & \multicolumn{1}{|l}{} & \multicolumn{1}{|l}{} &
\multicolumn{1}{|l}{} & \multicolumn{1}{|l}{} &
\multicolumn{1}{|l|}{} & $c_{3}=0.03$ \\ \hline\hline $^{178}Hf$ &
7 & \multicolumn{1}{|l}{0} & \multicolumn{1}{|l}{0} &
\multicolumn{1}{|l}{0} & \multicolumn{1}{|l}{0} &
\multicolumn{1}{|l|}{0.0010 } &
\begin{tabular}{l}
$a=0.02666$ \\
$b=-0.00007$%
\end{tabular}
\\
& 9 & \multicolumn{1}{|l}{2} & \multicolumn{1}{|l}{4} &
\multicolumn{1}{|l}{2 } & \multicolumn{1}{|l}{0} &
\multicolumn{1}{|l|}{0.0008} & $\alpha
_{3}=0.02677$ \\
& 9 & \multicolumn{1}{|l}{4} & \multicolumn{1}{|l}{4} &
\multicolumn{1}{|l}{2 } & \multicolumn{1}{|l}{2} &
\multicolumn{1}{|l|}{0.0016} & $\alpha
_{1}=0.05274$ \\
& 7 & \multicolumn{1}{|l}{6} & \multicolumn{1}{|l}{6} &
\multicolumn{1}{|l}{ 2} & \multicolumn{1}{|l}{2} &
\multicolumn{1}{|l|}{0.0008} & $\beta _{3}=0.01482$ \\
\hline\hline $^{154}Gd$ & 11 & \multicolumn{1}{|l}{0} &
\multicolumn{1}{|l}{0} & \multicolumn{1}{|l}{0} &
\multicolumn{1}{|l}{0} & \multicolumn{1}{|l|}{0.0023 } &
\begin{tabular}{l}
$a=0.05218$ \\
$b=-0.00019$%
\end{tabular}
\\
& 17 & \multicolumn{1}{|l}{2} & \multicolumn{1}{|l}{4} &
\multicolumn{1}{|l}{ 2} & \multicolumn{1}{|l}{0} &
\multicolumn{1}{|l|}{0.0482} & $\alpha
_{3}=0.0400$ \\
& 8 & \multicolumn{1}{|l}{4} & \multicolumn{1}{|l}{4} &
\multicolumn{1}{|l}{2 } & \multicolumn{1}{|l}{2} &
\multicolumn{1}{|l|}{0.0027} & $\alpha
_{1}=0.09574$ \\
& 6 & \multicolumn{1}{|l}{6} & \multicolumn{1}{|l}{8} &
\multicolumn{1}{|l}{4 } & \multicolumn{1}{|l}{2} &
\multicolumn{1}{|l|}{0.0023} & $\beta
_{3}=0.01634$ \\
& 3 & \multicolumn{1}{|l}{3} & \multicolumn{1}{|l}{4} &
\multicolumn{1}{|l}{2
} & \multicolumn{1}{|l}{0} & \multicolumn{1}{|l|}{0.0033} & $c_{3}=0.05$ \\
& 2 & \multicolumn{1}{|l}{5} & \multicolumn{1}{|l}{6} &
\multicolumn{1}{|l}{2
} & \multicolumn{1}{|l}{0} & \multicolumn{1}{|l|}{0.00008} & $c_{5}=0.09$ \\
\hline\hline
\end{tabular}
\label{coef}
\end{table}

As already mentioned, the examples chosen for the present study
consists of nuclei for which there is experimental data on
low-lying energies of more than five states in each of the angular
momenta $ L=0, 2, 4$ classes. Of the examples considered, two of
the nuclei have typical vibrational spectra \cite{Vibr}:
$^{144}Nd$ and $^{148}Sm$ while the rest $^{154}Gd,^{168}Er$ and
$^{178}Hf$ possess typical rotational character. This is confirmed
by the values obtained for the inertia parameter, $\beta _{3}$,
given in TABLE \ref{coef}. One of the main distinctions of these
two types of spectra is the position of the first excited
$2_{1}^{+}$ state above the GSB. For vibrational nuclei the number
is rather high, around $1MeV$,
    while for the well deformed nuclei it lies at about $0.07MeV$, roughly an
order of magnitude less than for the vibrational case.

For $^{144}Nd$ and $^{148}Sm$ which have vibrational spectra we
apply the procedure as described above with even $T$ values that
differ considerably ($\Delta T=4$) for the distinct sets of $L$
values. This corresponds to rather large changes in the values of
the initial $N_{min}=2T$. Most of the states with fixed $L$ are on
the left-hand-side of the symmetric parabolas, so the values of
$N_{L_{i}}$ increase with an increase in the energy of these
states. The step with which $N_{L_{i}}$ increases depends on the
energy differences between the states on the parabola. These
differences are the largest for the $0^{+}$ states and usually
decrease with increasing $L$. This is because all of the $0^{+}$
states are bandheads, and some of the $2^{+}$ and $4^{+}$ states
belong to these bands. By looking at the states as distributed on
the parabolas one can recognize their ordering into different
bands. The GSB is formed from the lowest states with
$J^{\pi}=0^{+},2^{+},4^{+}$ is almost equidistant in the case of
vibrational nuclei with very close or even equal values of $n$ for
the states in the other excited bands. The $ 0^{+},2^{+},4^{+}$
triplets of states that is almost degenerate in energy is
characteristic of the harmonic quadrupole vibrations that can be
observed on the theoretical energy curves, and are characterized
with almost equal differences between their respective values of
$N$. A good example is observed in FIG. \ref{NdSm} (left) for the
$^{144}Nd$ spectra, where such a triplet is formed by the $
0_{2}^{+},2_{3}^{+},4_{2}^{+}$ states with $n=25,20,15$,
respectively.

The other three examples, $^{154}Gd,^{168}Er$ and $^{178}Hf$,
shown in FIGs. \ref{ErHf} and \ref{Gd}, are typical rotational
nuclei. The rotational character of the spectra requires small
differences in the values of $N$ for states with $J^{\pi
}=0_{1}^{+},2_{1}^{+},4_{1}^{+}$ belonging to the GSB.  So, for
the sequences of $L$ states being considered we have to use almost
equal values  of $T$, which corresponds to nearly equal values of
$N_{\min }$.  To avoid a degeneracy of the energies with respect
to $N_{L_{i}}$ for the nuclei with rotational spectra, and to
clearly distinguish the parabolas, we have to use the symmetric
feature of the second order curves. This motivates our use of the
second solution $N_{0_{1}}^{\prime \prime }=-\frac{b}{a}$ of the
equation for the ground state (\ref{0energ}), defining the maximum
$ N_{0_{1}}$ for the ground state which yields a restriction on
the values of $N_{L_{i}}$ \cite{DraRo}. These increase for the
$L>0=2,4,6...$ states but not as much as in the vibrational case.
Hence, it is convenient in this case to place the states with a
given $L$ in the rotational spectra on the right-hand-side of the
theoretical curves. As a result, with increasing energy on a
specified parabola the number of bosons that are required to build
states decreases. In short, if the number of quanta required for a
collective state is taken as a measure of collectivity, the states
from a rotational spectra are much more collective than
vibrational ones, an expected result. In this case, one can also
observe the structure of collective bands that are formed by sets
of states from different curves. The best examples are for the GSB
and the first excited $\beta$- and $\gamma$-bands (see FIGs.
\ref{ErHf}-\ref{Gd}). In FIG. \ref{Gd}, for the spectra of
$^{154}Gd$, in addition to the $J^{\pi }=0^{+},2^{+},4^{+}$ states
we have included states with $J^{\pi }=3^{+},5^{+},6^{+}$. This
shows that the method also works for the $K^{\pi }=2^{+}$ and
$4^{+}$ bands. The respective values of $c_{L}$ are given in TABLE
\ref{coef}. In this case the second $0_{2}^{+}$ band is below the
$\gamma$-band in contrast with the $^{168}Er$  and $^{178}Hf$
spectra \cite{Hf}. The examples presented include a lot of
collective states, which cannot always be clearly distinguished on
the parabolas as they have almost equal values of $N$. For such
cases the symmetry feature of the second order curves can be used
to place some of them at the other side of the parabola in FIGs.
\ref{ErHf}-\ref{Gd}.

\section{\textbf{Conclusions}}

In this paper we introduced a theoretical framework for understanding
the empirical observation that collective states fall on two
parametric second order curves with respect to a variable $n$ that
counts the number of collective phonons (bosons) that is used to
build the corresponding many-particle configurations.

The theory is based on the reduction of $Sp(12,R)$ into to the
direct product $Sp(4,R)\otimes SO(3)$. The overarching $Sp(12,R)$
structure is the dynamical symmetry group of the Interacting
Vector Boson Model (IVBM), a phenomenological model that has been
shown to be successful in a description of collective nuclear
states. The $SO(3)$ angular momentum group, through its role which
is complementary to $Sp(4,R)$, labels states with a fixed angular
momentum $L$. The $Sp(4,R)$ basis is obtained via a reduction of
its boson representations into irreps of $SU(2)$ which are labeled
by the quantum numbers $ T,T_{0}$ of the pseudo-spin and its third
projection. The first-order invariant of the $U(2)\subset Sp(4,R)$
is the total number of vector bosons $ N$ of the IVBM. The
reduction that one exploits follows from the reduction of
$Sp(12,R)$ to its maximal compact subgroup $U(6)\supset
U(2)\otimes SU(3)$, which gives the rotational limit of the IVBM.

As a consequence, we use the model Hamiltonian of this limit and its
eigenstates to
obtain the theoretical energies of the sets of states with fixed
angular momentum
$L$. These fall on parabolas with respect to the variable $N=4n$. The
parameters of
the Hamiltonian are evaluated through a fitting procedure to known experimental
energies of the sets of states under consideration, that is, those with $J^{\pi
}=0^{+},2^{+},4^{+}$, and the appropriate value of $N$ for each
state. The examples
presented show that the procedure is accurate and appropriate for typical
collective states that are observed in atomic nuclei. In short, the
theory can be
applied to typical collective vibrational and rotational spectra and
can be used to
clearly distinguished these using symmetry properties of the second
order curves. The
vibrational nuclei are placed on the left-hand-side of the parabolas with
differing values of $T$ and with $N$ increasing with increasing
energies. In the
rotational case the situation is the opposite, which confirms the
traditional view
of vibrational states as few phonon states and rotational ones that involve a
higher level of collective coherence, namely, more bosons. The band
structure and
the energy degeneracies in both cases are also clearly observed.

The results introduced here to illustrate the theory demonstrate that the IVBM
can be used to reproduce reliably empirical observations of the energy
distribution of collective states. Such a demonstration can be provided for any
collective model that includes one- and two-body interactions in the
Hamiltonian.
The main feature that leads to our parameterization is the symplectic dynamical
symmetry of the IVBM. This allows for a change in the number of
``phonons'' that
are required to build the states. This investigation also provides insight
into the structure of collective states, revealing the similar origin of
vibrational and rotational spectra, but at the same time yielding
information about unique features that distinguish the two cases.

\begin{acknowledgements}
Support from the U.S. National Science Foundation, Grant No. PHY-0140300,
is acknowledged. The authors are grateful to Dr. C. Bahri, Prof. D. Rowe
and N. V. Zamfir for the helpful and enlightening discussions.
\end{acknowledgements}

\end{document}